\documentclass[twocolumn,showpacs,preprintnumbers,amsmath,amssymb]{revtex4}
\usepackage{graphicx}
\usepackage{dcolumn}
\usepackage{bm}
\usepackage{float}
\usepackage{mathrsfs}
\usepackage{appendix}
\DeclareMathOperator{\sech}{sech}
\begin{document}

\title{Effect of coherence of nonthermal reservoirs on heat transport in a microscopic collision model}

\author{Lei Li$^{1,2}$}
\author{Jian Zou$^{1}$}
\email{zoujian@bit.edu.cn}
\author{Hai Li$^{3}$}
\author{Bao-Ming Xu$^{4}$}
\author{Yuan-Mei Wang$^{1}$}
\author{Bin Shao$^{1}$}

\affiliation{$^{1}$School of Physics, Beijing Institute of Technology, Beijing 100081, China}
\affiliation{$^{2}$Southwest Institute of Technical Physics, Chengdu, Sichuan 610041, P.R. China}
\affiliation{$^{3}$School of Information and Electronic Engineering, Shandong Technology and Business University, Yantai 264000, P.R. China}
\affiliation{$^{4}$School of Physics, Qufu Normal University, Qufu 273165, China}

\date{Submitted ****}

\begin{abstract}
We investigate the heat transport between two nonthermal reservoirs
based on a microscopic collision model. We consider a bipartite
system consisting of two identical subsystems, and each subsystem
interacts with its own local reservoir, which consists of a large
collection of initially uncorrelated ancillas. Then a heat transport
is formed between two reservoirs by a sequence of pairwise
collisions (inter-subsystem and subsystem-local reservoir). In this
paper we consider two kinds of reservoir's initial states, the
thermal state, and the state with coherence whose diagonal elements
are the same as that of the thermal state and the off-diaganal
elements are nonzero. In this way, we define the effective
temperature of the reservoir with coherence according to its
diagonal elements. We find that for two reservoirs having coherence
the direction of the steady current of heat is different for
different phase differences between the two initial states of two
reservoirs, especially the heat can transfer from the ``cold
reservoir'' to the ``hot reservoir'' in the steady regime for
particular phase difference. And in the limit of the effective
temperature difference between the two reservoirs $\Delta
T\rightarrow0$, for most of the phase differences, the steady heat
current increases with the increase of effective
temperature until to the high effective temperature limit; while for
the thermal state or particular phase difference the steady heat
current decreases with the increase of temperature at high
temperatures, and in this case the conductance can be obtained.
\end{abstract}


\maketitle

\section{Introduction}\label{Sec1}
The increasing abilities to control systems at smaller and smaller
scales motivate us to understand how the laws of physics developed
in the macroscopic domain are modified in the microscopic
scale~\cite{e2}. From the fundamental perspective, a link between
quantum dynamics and thermodynamic processes has been widely
investigated~\cite{e3,e5,e6}. In particular, out-of-equilibrium
thermodynamics of quantum systems represents one of the most active
research areas in this field~\cite{e7,e8,e9,g6,e10}. And
understanding how energy transport can be controlled and efficiently
distributed has been identified as one of the crucial studies for
the development of quantum thermodynamics~\cite{l4,l5,l6}. In common
construction, heat flow through a quantum system is generated by
coupling it to two thermal reservoirs with different
temperatures~\cite{c12,c13}, and heat transport problem has
attracted considerable attention during the past
decades~\cite{l1,l2,e4,l7,l3}. For example, Wichterich~\emph{et al.}
have investigated heat transport in a spin-1/2 Heisenberg chain,
locally coupled to independent thermal baths of different
temperatures, and they have obtained a stationary energy current by
quantum master equation method and provided the way for efficient
numerical investigations of heat transport in larger
systems~\cite{l1}; Werlang \emph{et al.} have realized a heat
transport between two pure-dephasing Markovian reservoirs connected
through a chain of coupled sites, and quantum coherence between
sites is generated in the steady regime and results in the
underlying mechanism sustaining the effect of heat
transport~\cite{l3}. Moreover, one of the conceptual pillars in
energy transport, Fourier's law of heat conduction, has become an
important issue and has been investigated in classical~\cite{d6,d7}
and quantum systems~\cite{e4,d8,d9,d11}. An important step in
understanding how Fourier's law emerges from the quantum domain has
been made by Michel~\emph{et al.}~\cite{d8}. Subsequently
Manzano~\emph{et al.} have analyzed the steady energy transfer in a
chain of coupled two-level systems connecting two thermal
reservoirs, and they have revealed that there is a distinct
violation of Fourier's law in the quantum transport scenario of
their model~\cite{e4}.

As one of the representative model for studying open quantum system,
collision model has been extensively studied during the past
decades~\cite{d12,d13,d14,d15,d16,d17,d18,d19,d20,d21,d22,d23,d24}.
A quantum collision model~\cite{d20,d25} is a microscopic framework
to describe the open dynamics of a system interacting with a
reservoir assumed to be consisted of a large collection of smaller
constituents (ancillas). The system is assumed to interact with the
environment via a sequence of ``collisions'' between the system and
ancillas, and each collision being described by the same bipartite
quantum map (usually a unitary one). The reduced system can be
obtained in many cases without any approximations, hence the
complete positivity (CP) of the dynamical map is guaranteed. The
collision model has been applied in quantum thermodynamics recently.
For example, Lorenzo \emph{et al.} have investigated the link
between information and thermodynamics in the dynamics of a
multipartite open quantum system, which is described in terms of a
collision model with a finite temperature reservoir~\cite{d18}; and
they have also explored the relation between heat flux and quantum
correlations of a bipartite system via a collision-model-based
approach~\cite{d19}. Moreover, Pezzutto \emph{et al.} have studied
the heat exchange between system and environment and the
information-to-energy conversion through a collision-based
model~\cite{d22}. Actually a similar model of a system repeatedly in
contact with a stream of independently prepared units being in any
nonequilibrium state (acting as a reservoir), has been investigated
from the perspective of quantum and information thermodynamics
~\cite{x2}. This kind of setups have been used in quantum optics,
theoretically as well as experimentally, for example, a micromaser
cavity interacts sequentially with a stream of flying atoms prepared
in nonequilibrium states~\cite{ji,j2,j3}, and this setup has also
been used to investigate the work extraction from nonequilibrium
bath including the particle units crossing the cavity each time
being single multi-level atom~\cite{n2}, single two-level atom~\cite{n3},
two- and three-atom clusters~\cite{n4,n5}.

As mentioned above, traditional thermodynamic setups consist of a system in contact with thermal reservoirs.
In fact, quantum systems also open up the possibility for exploring more general reservoirs~\cite{x1,x2}. Especially,
the effect of quantum nonequilibrium baths, for example reservoirs with quantum correlations~\cite{n1}, coherence~\cite{n2,n3,n4,n5}
and squeezing~\cite{n6,n7,n8} have been studied extensively in quantum thermodynamics. Recently, the exploration of the effect of a
reservoir's coherence in quantum thermodynamics has provoked great interest. Scully~\emph{et al.} firstly demonstrated that the improvement
of the working efficiency can be realized by using the nonequilibrium state preparation, i.e., the bath with coherence, and the obtained
efficiency is beyond the classical Carnot efficiency~\cite{n2}. This observation has been used in different scenarios~\cite{h2,h4}.
However we have not seen any report about the effect of coherence of reservoir on heat transport between two reservoirs.
Thus an interesting question concerns if, and how, the heat transfer between two reservoirs can be affected by intra-reservoir coherence.

\begin{figure}[h]
\includegraphics[scale=0.35]{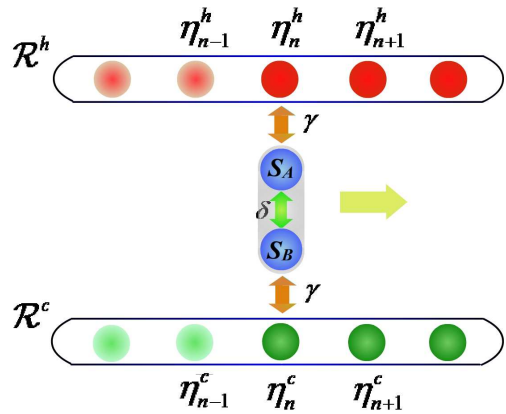}
\centering
\caption{(Color online) Sketch of the protocol: A bipartite system $\mathcal{S}$ consists of subsystems $\mathcal{S}_{A}$ and $\mathcal{S}_{B}$; after inter-subsystem interaction, $\mathcal{S}_{A}$ interacts with $n$th ancilla (prepared in $\eta^{h}_{n}$), and $\mathcal{S}_{B}$ interacts with $n$th ancilla (prepared in $\eta^{c}_{n}$); then this process is repeated and is directed to the $(n+1)$th ancilla. Thus the heat transport is established between the two reservoirs by a sequence of this process.}
\label{Fig1}
\end{figure}

In this paper, we consider a bipartite system which, interacts with its local reservoir consisting of a large collection of initially uncorrelated systems which we call ancillas, respectively (see Fig. 1). We consider two kinds of reservoir's initial states, one is the thermal state and the other is the state with coherence. The diagonal elements of both states are identical, and compared with the thermal state, the off-diagonal elements of the state with coherence are nonzero. In this way we define the effective temperature of the reservoir with coherence according to its diagonal elements. Based on this, the heat transport is formed between two reservoirs of different effective temperature by a sequence of pairwise collisions (inter-subsystem and subsystem-local reservoir). If the ancillas of both reservoirs are in thermal states, the general form of heat transport appears, i.e., heat transfers from the hot reservoir to the cold reservoir in the steady regime. However in the case of ancillas' states with coherence, the direction of steady heat current can be changed by manipulating the relative phase of ancillas. And in the limit of the effective temperature difference between the two reservoirs $\Delta T\rightarrow0$, for most of phase differences the steady heat current increases with the increase of effective temperature until to the high effective temperature limit. While for the thermal state or particular phase difference ($0$ and $\pi$) the steady heat current decreases with the increase of temperature at high temperatures, and in this case the conductance can be obtained.

\section{Model}\label{Sec2}
We consider a bipartite system $\mathcal{S}$, consisting of two identical subsystem: qubits $\mathcal{S}_{A}$ and $\mathcal{S}_{B}$,
with logical states $\{|0\rangle,|1\rangle\}$ for each qubit. Subsystem $\mathcal{S}_{A}$ is in contact with a reservoir $\mathcal{R}^{h}$ consisted of a collection of $N$ identical noninteracting ancillas $\{\mathcal{R}^{h}_{1},\mathcal{R}^{h}_{2},...,\mathcal{R}^{h}_{N}\}$, and subsystem $\mathcal{S}_{B}$ is also in contact with a reservoir $\mathcal{R}^{c}$ consisted of a collection of $N$ identical noninteracting ancillas $\{\mathcal{R}^{c}_{1},\mathcal{R}^{c}_{2},...,\mathcal{R}^{c}_{N}\}$. Here we focus on the ancillas of the reservoir $\mathcal{R}^{h}$ ($\mathcal{R}^{c}$) consisting of simple two-level systems (qubits) whose logical states are $\{|0\rangle,|1\rangle\}_{j} (j=1,...,N)$. The two reservoirs are, therefore, in the product state $\eta_{tot}^{h}=\otimes^{N}_{j=1}\eta^{h}_{j}$ and $\eta_{tot}^{c}=\otimes^{N}_{j=1}\eta^{c}_{j}$ respectively. The evolution of the whole system including the two subsystems and their reservoirs is: $\mathcal{S}_{A}$ and $\mathcal{S}_{B}$ interact first, subsequently $\mathcal{S}_{A}$ and $\mathcal{S}_{B}$ collide with the individual ancillas of their local reservoirs, respectively; and then this process is repeated. The general scheme is illustrated in Fig. 1. The assumption of two big reservoirs ($\mathcal{R}^{h}$ and $\mathcal{R}^{c}$) implies that the subsystem never interacts twice with the same ancilla, i.e., at each collision the state of the ancilla is refreshed.

In our model, we consider a coherent interaction between the subsystem and the $j$th ancilla of the corresponding reservoir, i.e., a mechanism that can be described by a Hamiltonian model of some form, specifically in this paper we suppose that the interaction Hamiltonian is
\begin{equation}
\hat{H}_{int}=g(\hat{\sigma}^{\mathcal{S}_{A(B)}}_{x}\hat{\sigma}^{\mathcal{R}^{h(c)}_{j}}_{x}+\hat{\sigma}^{\mathcal{S}_{A(B)}}_{y}\hat{\sigma}^{\mathcal{R}^{h(c)}_{j}}_{y}+\hat{\sigma}^{\mathcal{S}_{A(B)}}_{z}\hat{\sigma}^{\mathcal{R}^{h(c)}_{j}}_{z}),
\end{equation}
where $\hat{\sigma}^{\mathcal{S}_{A(B)}}_{i}$ and $\hat{\sigma}^{\mathcal{R}^{h(c)}_{j}}_{i}$ $(i=x,y,z)$ are the Pauli matrices, and $g$ is a coupling constant; and each collision is described by a unitary operator $\hat{U}_{\mathcal{S}_{A(B)},\mathcal{R}^{h(c)}_{j}}=e^{-i\hat{H}_{int}\tau}$, where $\tau$ is the collision time and we set $\hbar=1$ throughout this paper. Using the following result~\cite{d12}
\begin{equation}
e^{i\frac{\phi}{2}(\hat{\sigma}_{x}\otimes\hat{\sigma}_{x}+\hat{\sigma}_{y}\otimes\hat{\sigma}_{y}+\hat{\sigma}_{z}\otimes\hat{\sigma}_{z})}=e^{-i\frac{\phi}{2}}(\cos\phi\hat{\mathbb{I}}+i\sin\phi\hat{S}^{sw}),
\end{equation}
where $\hat{\mathbb{I}}$ is the identity operator, and $\hat{S}^{sw}$ is the two-particle swap operator, i.e., it is the unitary operation whose action is $|\psi_{1}\rangle\otimes|\psi_{2}\rangle\rightarrow|\psi_{2}\rangle\otimes|\psi_{1}\rangle$ for all $|\psi_{1}\rangle,|\psi_{2}\rangle\in\mathbb{C}^{2}$. We can now write the unitary time-evolution operator
\begin{equation}
\hat{U}_{\mathcal{S}_{A(B)},\mathcal{R}^{h(c)}_{j}}=(\cos\gamma)\hat{\mathbb{I}}_{\mathcal{S}_{A(B)},\mathcal{R}^{h(c)}_{j}}+i(\sin\gamma)\hat{S}_{\mathcal{S}_{A(B)},\mathcal{R}^{h(c)}_{j}}^{sw},
\end{equation}
where $\gamma=2g\tau$ is a dimensionless interaction strength. And we have assumed above that each ancilla of the two reservoirs have two energy levels, and in the ordered basis $\{|00\rangle, |01\rangle, |10\rangle, |11\rangle\}$, $\hat{S}_{\mathcal{S}_{A(B)},\mathcal{R}^{h(c)}_{j}}^{sw}$ in Eq. (3) reads~\cite{a3}
\begin{equation}
\hat{S}_{\mathcal{S}_{A(B)},\mathcal{R}^{h(c)}_{j}}^{sw}=
\begin{pmatrix}
1& \hspace{0.25cm}0 & \hspace{0.25cm}0 & \hspace{0.25cm}0\\
\*0 &\hspace{0.25cm}0 & \hspace{0.25cm}1& \hspace{0.25cm}0\\
\*0 &\hspace{0.25cm}1 & \hspace{0.25cm}0& \hspace{0.25cm}0\\
0& \hspace{0.25cm}0 & \hspace{0.25cm}0 & \hspace{0.25cm}1%
\end{pmatrix}%
\label{initial}.
\end{equation}
Similar to Eq. (3), the interaction between two subsystems $\mathcal{S}_{A}$ and $\mathcal{S}_{B}$ is implemented through the unitary evolution
\begin{equation}
\hat{V}_{\mathcal{S}_{A},\mathcal{S}_{B}}=(\cos\delta)\hat{\mathbb{I}}_{\mathcal{S}_{A},\mathcal{S}_{B}}+i(\sin\delta)\hat{S}_{\mathcal{S}_{A},\mathcal{S}_{B}}^{sw},
\end{equation}
with $\delta\neq\gamma$, in general, and the analog of the operations introduced above applies to $\hat{\mathbb{I}}_{S_{A},S_{B}}$, $\hat{S}^{sw}_{\mathcal{S}_{A},\mathcal{S}_{B}}$ (swap gate between two subsystems). Such interactions give rise to the dynamical maps
\begin{equation}
\hat{\Phi}_{\mathcal{S}_{A(B)},\mathcal{R}^{h(c)}_{j}}[\rho]=\hat{U}_{\mathcal{S}_{A(B)},\mathcal{R}^{h(c)}_{j}}\rho\hat{U}^{\dagger}_{\mathcal{S}_{A(B)},\mathcal{R}^{h(c)}_{j}},
\end{equation}

\begin{equation}
\hat{\Psi}_{\mathcal{S}_{A},\mathcal{S}_{B}}[\rho]=\hat{V}_{\mathcal{S}_{A},\mathcal{S}_{B}}\rho\hat{V}^{\dagger}_{\mathcal{S}_{A},\mathcal{S}_{B}}.
\end{equation}

As mentioned above the dynamics of system $\mathcal{S}$ consists of sequential inter-system interaction interspersed with subsystem-reservoir collisions. Therefore, each collision is treated in the following process, specifically $\mathcal{S}_{A}$ and $\mathcal{S}_{B}$ interact first, subsequently $\mathcal{S}_{A}$ and $\mathcal{S}_{B}$ collide with one of the ancillas in $\mathcal{R}^{h}$ and $\mathcal{R}^{c}$ respectively. Thus the system is brought from step $n$ to step $n+1$ through the process
\begin{equation}
\rho^{\mathcal{S}}_{n}\otimes\eta_{n+1}^{h}\otimes\eta_{n+1}^{c}\rightarrow \rho^{\mathcal{S}E}_{n+1}=\hat{\mathcal{U}}(\rho^{\mathcal{S}}_{n}\otimes\eta_{n+1}^{h}\otimes\eta_{n+1}^{c})\hat{\mathcal{U}}^{\dagger},
\end{equation}
with $\hat{\mathcal{U}}$ the overall unitary evolution experienced by the $SE$ system (system plus the $(n+1)$th ancillas $\mathcal{R}^{h}_{n+1}$, $\mathcal{R}^{c}_{n+1}$) being generated by the composition of the set of unitary gates introduced above, i.e., $\hat{\mathcal{U}}=\hat{U}_{\mathcal{S}_{B},\mathcal{R}^{c}_{n+1}}(\gamma)\hat{U}_{\mathcal{S}_{A},\mathcal{R}^{h}_{n+1}}(\gamma)\hat{V}_{\mathcal{S}_{A},\mathcal{S}_{B}}(\delta)$. Hence after the $(n+1)$th collision, the state of system is $\rho^{\mathcal{S}}_{n+1}=\mathrm{Tr}_{\mathcal{R}^{h}_{n+1}\mathcal{R}^{c}_{n+1}}[\rho^{\mathcal{S}E}_{n+1}]$, and the state of the $(n+1)$th ancilla of reservoir $\mathcal{R}^{h}$ ($\mathcal{R}^{c}$)
\begin{equation}
\tilde{\eta}^{h(c)}_{n+1}=\mathrm{Tr}_{\mathcal{S}\mathcal{R}^{c(h)}_{n+1}}[\rho^{\mathcal{S}E}_{n+1}],
\end{equation}
where $\mathrm{Tr}_{\mathcal{S}\mathcal{R}^{c(h)}_{n+1}}$ means the trace of $\mathcal{S}$ and $\mathcal{R}^{h(c)}_{n+1}$ degrees of freedom. And note that in Appendix A, we have demonstrated that the total change of energy of reservoir $\mathcal{R}^{h(c)}$ after the $(n+1)$th collision, is equivalent to the sum of energy change of the $j$th ancilla (in reservoir $\mathcal{R}^{h(c)}$) during the $j$th collision. Therefore from Eq. (9) we can obtain the heat exchange between reservoir $\mathcal{R}^{h(c)}$ and system during the $(n+1)$th collision
\begin{equation}
\Delta\mathrm{Q}^{\mathcal{R}^{h(c)}}_{n+1}=\mathrm{Tr}[\hat{H}^{h(c)}_{n+1}(\tilde{\eta}^{h(c)}_{n+1}-\eta^{h(c)}_{n+1})],
\end{equation}
where $\hat{H}^{h(c)}_{n+1}=\frac{1}{2}\omega\hat{\sigma}_{z}$ (in this paper we assume that $\hat{H}^{h(c)}_{j}=\frac{1}{2}\omega\hat{\sigma}_{z}$ is the local Hamiltonian of each ancilla of two reservoirs). And we consider energy-conserving $\mathcal{S}_{A(B)}-\mathcal{R}^{h(c)}_{j}$ interactions, i.e., $[\hat{H}_{int}, (\hat{H}_{\mathcal{S}_{A(B)}}+\hat{H}^{h(c)}_{j})]=0$ (this can be realized in the case of resonance interaction between $\mathcal{S}_{A(B)}$ and the ancilla). Hence, $\Delta\mathrm{Q}^{\mathcal{R}^{h}}_{n+1}=-\Delta\mathrm{Q}^{\mathcal{R}^{c}}_{n+1}$ in the steady regime. Therefore, Eq. (10) can also be defined as the heat current
\begin{equation}
J_{h(c)}=\Delta\mathrm{Q}^{\mathcal{R}^{h(c)}}_{n+1},
\end{equation}
i.e., the heat flows into (or out of) reservoir $\mathcal{R}^{h(c)}$ during the $(n+1)$th collision.

In order to investigate the effect of coherence of reservoirs on the heat current, we consider the initial state of each ancilla of two reservoirs as
\begin{equation}
\rho_{coh}=p|\psi\rangle\langle\psi|+(1-p)\rho_{\beta},
\end{equation}
where $p\in[0,1]$, $|\psi\rangle=\frac{1}{\sqrt{Z}}(e^{-\frac{1}{4}\omega\beta}|0\rangle+e^{i\phi+\frac{1}{4}\omega\beta}|1\rangle)$ with a relative phase $\phi$, and $\rho_{\beta}$ is the thermal state assumed to be of canonical equilibrium form, i.e.,  $\rho_{\beta}=\frac{1}{Z}e^{-\beta\hat{H}^{h(c)}_{j}}$. Here $\beta=1/T_{h(c)}$ and $Z=\textmd{Tr}[e^{-\beta\hat{H}^{h(c)}_{j}}]$ are the corresponding inverse temperature and the partition function. Note that the diagonal elements of states $\rho_{coh}$ and $\rho_{\beta}$ are identical, and compared with the thermal state, the off-diagonal elements of state $\rho_{coh}$ are nonzero if $p\neq0$. Therefore, Eq. (12) can also be written as
\begin{equation}
\rho_{coh}=\rho_{\beta}+p\rho_{non},
\end{equation}
where $\rho_{non}$ is the non-diagonal part of state
$|\psi\rangle\langle\psi|$, i.e., the off-diagonal elements of
$\rho_{non}$ are the same as that of state
$|\psi\rangle\langle\psi|$ and the diagonal elements are zero. As
the diagonal elements of state $\rho_{coh}$ are the same as that of
the thermal state, the effective temperature of the reservoir is defined by
its diagonal elements. In this paper, we assume that the ancillas of reservoir
$\mathcal{R}^{h}$ and $\mathcal{R}^{c}$ are in state (13) with different effective
temperature $T_{h}$ and $T_{c}$ respectively, and $T_{h}>T_{c}$, which composes the ``hot
reservoir'' $\mathcal{R}^{h}$ and the ``cold reservoir''
$\mathcal{R}^{c}$. Based on this, an interesting question concerns
if, and how, the heat current can be affected by the non-zero
off-diagonal elements of state $\rho_{coh}$, i.e.,
reservoir-coherence.

\section{Heat current}\label{Sec3}
We first consider the heat flow from the reservoir
$\mathcal{R}^{h}$, i.e., $J_{h}$. In Fig. 2, we plot the heat
current $J_{h}$ against the number of collisions $n$. We suppose
that the two reservoirs are in the state (13) and the system is
initially in the ground state
$|\psi\rangle_{\mathcal{S}}=|11\rangle$. And we let $T_{h}=2$ and
$T_{c}=1$ in Eq. (13) for reservoirs $\mathcal{R}^{h}$ and
$\mathcal{R}^{c}$ respectively. Based on this we consider two
initial states of the two reservoirs, one is $p=0.8$ with different
phase difference between the two reservoirs
$\phi_{h}-\phi_{c}=\{0,\pi/4,\pi/2,\pi,5\pi/4,3\pi/2\}$ (the two
reservoirs with coherence); and the other is $p=0$, i.e., the
thermal states $\rho_{\beta}$ of two reservoirs. It can be seen from
Fig. 2 that a unidirectional heat current (heat flows out of hot
reservoir $\mathcal{R}^{h}$) is formed during the whole dynamics for
the two reservoirs with thermal state (dashed black line), which is
in contrast to the case of two reservoirs with coherence. For the
two reservoirs with coherence, from numerical calculations we find
that the heat current $J_{h}$ is independent of their respective
phases of two reservoirs, and is dependent only on the phase
difference between the two reservoirs, i.e., $\phi_{h}-\phi_{c}$.
And in this case, it can be seen from Fig. 2 that the heat currents
are oscillating between positive and negative values as $n$
increases, which reveals that the energy flows into (positive
$J_{h}$) and out of (negative $J_{h}$) reservoir $\mathcal{R}^{h}$
in the early stages of dynamics. And then at the later stage the
heat currents reach their stationary values respectively. And from
numerical calculations we find that the steady heat current of
reservoir $\mathcal{R}^{h}$ ($J^{steady}_{h}$) and the steady heat
current of reservoir $\mathcal{R}^{c}$ ($J^{steady}_{c}$) satisfy
the relation $J^{steady}_{h}=-J^{steady}_{c}$ as expected. When
$\phi_{h}-\phi_{c}\in[0,\pi]$, $J^{steady}_{h}$ are negative, which
corresponds to heat flowing from $\mathcal{R}^{h}$ to
$\mathcal{R}^{c}$ in the steady regime. However an interesting
thing appears: When $\phi_{h}-\phi_{c}\in(\pi,2\pi)$,
$J^{steady}_{h}$ are positive, which corresponds to heat flowing
from the ``cold reservoir'' ($\mathcal{R}^{c}$) to the ``hot
reservoir'' ($\mathcal{R}^{h}$), i.e., the ``cold reservoir'' is
`cooled'. In other words, we can realize a countercurrent of heat
from the ``cold reservoir'' to the ``hot one'' for particular phase
differences. This could appear to be counterintuitive at first, as
heat always transfers from hot reservoir to cold reservoir in
general. This can be understood as follows. First from Eqs.
(8)-(11), after some calculations we can obtain the steady heat
current (for simplicity we choose $\gamma=\pi/2$, i.e., a complete
swap between the subsystem and the corresponding ancilla) as
\begin{equation}
\begin{split}
J^{steady}_{h}=&J_{\beta}+J_{coh} \\
=&\frac{\sin^{2}\delta(e^{\beta_{h}}-e^{\beta_{c}})}{(1+e^{\beta_{h}})(1+e^{\beta_{c}})} \\
&-\frac{p^{2}\sin(2\delta)e^{\frac{\beta_{h}+\beta_{c}}{2}}}{(1+e^{\beta_{h}})(1+e^{\beta_{c}})}\sin(\phi_{h}-\phi_{c}),
\end{split}
\end{equation}
where $\delta\in(0,\frac{\pi}{2})$. Obviously, the steady heat
current can be divided into two parts, the first term in Eq. (14),
$J_{\beta}=\frac{\sin^{2}\delta(e^{\beta_{h}}-e^{\beta_{c}})}{(1+e^{\beta_{h}})(1+e^{\beta_{c}})}$
is the contribution of the diagonal elements in Eq. (13) and is
independent of $p$. In other words, $J_{\beta}$ is the steady
current of heat for the states of two reservoir without coherence,
i.e., the thermal states. And from the expression of $J_{\beta}$ it
can be seen that the direction of heat current depends on the
temperature difference between the two reservoirs. As in our paper
$T_{h}>T_{c}$, so $J_{\beta}<0$ is always satisfied. The second term
in Eq. (14),
$J_{coh}=-\frac{p^{2}\sin(2\delta)e^{\frac{\beta_{h}+\beta_{c}}{2}}\sin(\phi_{h}-\phi_{c})}{(1+e^{\beta_{h}})(1+e^{\beta_{c}})}$
is the contribution of non-diagonal elements in Eq. (13), i.e., the
contribution of coherence, and depends on the parameter $p$. Note
that Eq. (14) is the steady heat current in the case of complete swap,
and we analyze how the phase difference influence it in this case in
the following. From Eq. (14), if $\phi_{h}-\phi_{c}\in(0,\pi)$,
$J_{coh}<0$ and $J^{steady}_{h}<0$, which corresponds to heat
flowing from the ``hot reservoir'' to the ``cold reservoir''
($\mathcal{R}^{h}\longrightarrow\mathcal{R}^{c}$) in the
steady regime, and now $J_{h}^{steady}$ is larger than that of
the initial thermal state $\rho_{\beta}$. If
$\phi_{h}-\phi_{c}\in(\pi,2\pi)$, $J_{coh}>0$, the sign of
$J^{steady}_{h}$ depends on the absolute values of $J_{\beta}$ and
$J_{coh}$. In other words, in this case the direction of steady flow
of heat depends upon different contribution of the effective temperature
difference and coherence of reservoirs. For example when
$\phi_{h}-\phi_{c}=5\pi/4,3\pi/2$, $J_{\beta}<0$, $J_{coh}>0$,
and $|J_{\beta}|<|J_{coh}|$, which leads to $J^{steady}_{h}>0$ and a
heat current from the ``cold reservoir'' into the ``hot reservoir''
($\mathcal{R}^{c}\longrightarrow\mathcal{R}^{h}$) appears. And, if
$\phi_{h}-\phi_{c}=0,\pi$, $J_{coh}=0$ and
$J^{steady}_{h}=J_{\beta}$ which returns to the cases of initial
thermal state introduced above.

\begin{figure}[h]
\includegraphics[scale=0.58]{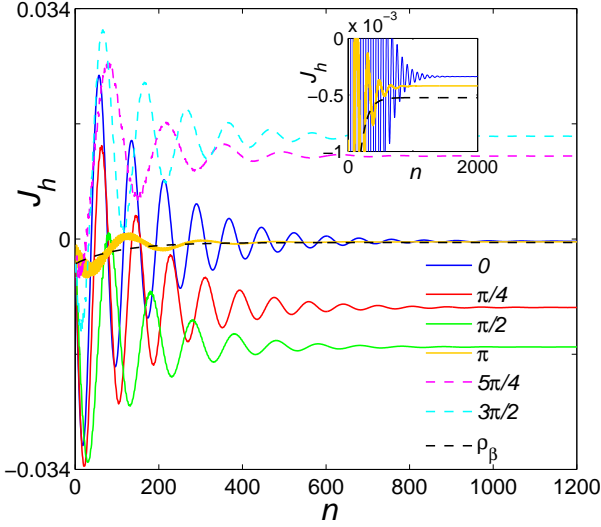}
\centering
\caption{(Color online) Heat current $J_{h}$ against the number of collisions $n$ for initial state (13) with $p=0.8$, and different relative phase $\phi_{h}-\phi_{c}=\{0,\pi/4,\pi/2,\pi,5\pi/4,3\pi/2\}$, and the dashed black line corresponds to $p=0$ in Eq. (13), i.e., the thermal state $\rho_{\beta}$. The inset is the magnified $J_{h}$ for the thermal state and state (13) with $\phi_{h}-\phi_{c}=0,\pi$. For all plots we choose the ground state of system $|\psi\rangle_{S}=|11\rangle$, $\gamma=\pi/32$, $\delta=\pi/4$, $T_{h}=2T_{c}=2$ (i.e., $\beta_{h}=\frac{1}{2}\beta_{c}=\frac{1}{2}$), and $\omega=1$.}
\label{Fig2}
\end{figure}

\begin{figure}[h]
\includegraphics[scale=0.58]{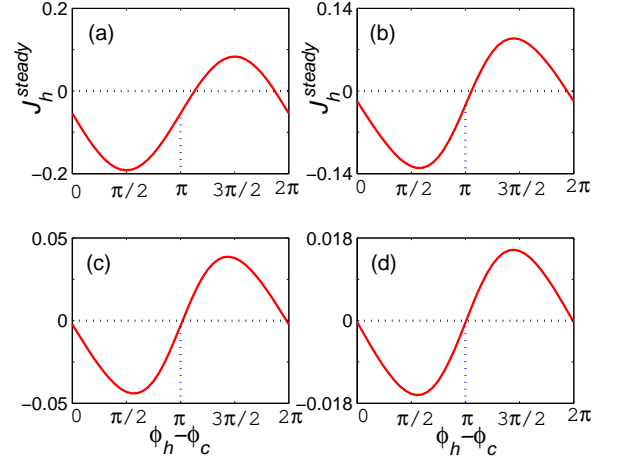}
\centering
\caption{(Color online) Steady heat current $J^{steady}_{h}$ as a function of $\phi_{h}-\phi_{c}$ for initial state (13) with $p=0.8$ and $\delta=\pi/4$, and different interaction strength (a) $\gamma=\pi/2$, (b) $\gamma=\pi/4$, (c) $\gamma=\pi/12$ and (d) $\gamma=\pi/32$. For all plots we choose the ground state of system $|\psi\rangle_{S}=|11\rangle$, $T_{h}=2T_{c}=2$ (i.e., $\beta_{h}=\frac{1}{2}\beta_{c}=\frac{1}{2}$), and $\omega=1$.}
\label{Fig3}
\end{figure}

Above, we have discussed the case of completely swap
($\gamma=\pi/2$), and now we analyze how the interaction strength
$\gamma$ influence the steady heat current $J^{steady}_{h}$. In Fig.
3, we plot $J^{steady}_{h}$ as a function of $\phi_{h}-\phi_{c}$ for
initial state (13) with $p=0.8$ and $\delta=\pi/4$, and different
$\gamma$, $\gamma=\{\pi/2,\pi/4,\pi/12,\pi/32\}$. It can be seen
from Fig. 3 that $J^{steady}_{h}$ moves down overall with the
increase of $\gamma$, which leads to the region of
$J^{steady}_{h}<0$ increasing and that of $J^{steady}_{h}>0$
decreasing, and the maximal difference between the two regions
corresponding to $J^{steady}_{h}>0$ and $J^{steady}_{h}<0$
respectively is reached when $\gamma=\pi/2$ (Fig. 3 (a)). It is noted that
$J_{\beta}<0$, therefore, it reveals that the effect of effective temperature
difference (i.e., $J_{\beta}$) on $J^{steady}_{h}$ is weakened with
the decrease of $\gamma$. In other words, with the decrease of
$\gamma$, the contribution of reservoir-coherence on
$J^{steady}_{h}$ is relatively enhanced. However, in practice the
system-bath coupling is weak in general, therefore we mainly
consider the interaction strength $\gamma=\pi/32$ in this paper. In
the case of $\gamma=\pi/32$, for $\phi_{h}-\phi_{c}=0,\pi$ the
steady heat current depends not only on the effective temperatures of two
reservoirs but also on the reservoir-coherence, i.e.,
$J^{steady}_{h}\neq J_{\beta}$ (in contrast to
$J^{steady}_{h}=J_{\beta}$ in the case of completely swap). This can
be seen from the inset of Fig. 2: For $\phi_{h}-\phi_{c}=0,\pi$
though the direction of the steady heat current is the same as that
of the initial thermal state, the magnitudes are slightly different,
which satisfy the relation
$J^{steady}_{h}(\phi_{h}-\phi_{c}=0)<J^{steady}_{h}(\phi_{h}-\phi_{c}=\pi)<J^{steady}_{h}(\beta)$.
Especially the heat currents in the early stages of dynamics (before
reaching the stationary state) are greatly different. When
$\phi_{h}-\phi_{c}=0,\pi$ the heat flows out of ``hot reservoir''
$\mathcal{R}^{h}$ and flows back into $\mathcal{R}^{h}$
alternatively before reaching the stationary state, and the backflow
of heat into $\mathcal{R}^{h}$ is especially obvious for
$\phi_{h}-\phi_{c}=0$, which is in contrast to the thermal state as
mentioned above (a unidirectional heat current during the whole
dynamics). In other words, even if there is no phase difference
between the two reservoirs, the heat current is also different from
that of the thermal state. Though the phase difference of two
reservoirs is zero, the two reservoirs with coherence can also
influence the heat transfer between the two reservoirs. And from the
discussion above, the heat transport strongly depends on the phase
difference between two reservoirs resulted from the quantum
interference effects of the two reservoirs.

From Eq. (13), we know that the amount of coherence of reservoir is dependent on the value of $p$, and the amount of coherence increase with the increase of $p$. Therefore we will investigate how the amount of reservoir-coherence influence the heat transfer
next. In the limit $p\rightarrow0$ the effect of reservoir-coherence on $J^{steady}_{h}$ is negligible, i.e., $J^{steady}_{h}\sim J_{\beta}$, thus the reversed steady current ($\mathcal{R}^{c}\rightarrow\mathcal{R}^{h}$) disappears. In the case of complete swap ($\gamma=\pi/2$): When $J^{steady}_{h}=0$ in Eq. (14) with $\phi_{h}-\phi_{c}\in(\pi,2\pi)$, we can obtain
\begin{equation}
p_{c}=\sqrt{\frac{\tan\delta(e^{\beta_{h}}-e^{\beta_{c}})}{2e^{\frac{\beta_{h}+\beta_{c}}{2}}\sin(\phi_{h}-\phi_{c})}},
\end{equation}
which is the critical value of $p$ appearing reversed steady current, i.e., we can not obtain the reversed steady current if $p<p_{c}$. When $\gamma\neq\pi/2$ (partial swap) it is difficult to obtain the analytical expression of the steady heat current, and with the same parameters in Fig. 2, $p\approx0.155$ is the critical value appearing the reversed steady current now. In other words, in this case we can realize the steady heat transfer from the ``cold reservoir'' ($T_{c}=1$) into the ``hot reservoir'' (${T_{h}}=2$) by manipulating the relative phase only when $p\gtrsim0.155$.

\section{Entropy change and heat current}\label{Sec4}
As mentioned above the evolution of the total system (system$+$two reservoirs) is unitary, therefore the entropy of the total system is conserved in this process. Based on this, and according to Ref.~\cite{c9}, for the initial separable state of the total system, $\rho^{tot}_{0}=\rho_{0}^{\mathcal{S}}\otimes\eta_{0}^{\mathcal{R}}$ ($\eta_{0}^{\mathcal{R}}=\eta_{tot}^{h}\otimes\eta_{tot}^{c}$), the change of the von Neuman entropy of system $\mathcal{S}$ after the $(n+1)$th collision, can be expressed as
\begin{equation}
\begin{split}
\Delta\mathrm{S}=&\mathrm{S}(\rho^{\mathcal{S}}_{n+1})-\mathrm{S}(\rho_{0}^{\mathcal{S}}) \\
=&D[\rho^{tot}_{n+1}\parallel\rho^{\mathcal{S}}_{n+1}\otimes^{n+1}_{j=1}\eta_{j}^{hc}]
+\sum^{n+1}_{j=1}\textrm{Tr}_{j}(\tilde{\eta}^{hc}_{j}-\eta^{hc}_{j})\ln\eta^{hc}_{j},
\end{split}
\end{equation}
where $\rho^{tot}_{n+1}$ is the total state of system $\mathcal{S}$ and the $(n+1)$ ancillas in each reservoir which have interacted with the system, $\eta^{hc}_{j}=\eta_{j}^{h}\otimes\eta_{j}^{c}$ is the initial state of the $j$th ancilla of the two reservoirs totally, and $\tilde{\eta}^{hc}_{j}=\textrm{Tr}_{\neq j}(\rho^{tot}_{n+1})$ is the marginal state of `$\mathcal{R}^{h}_{j}+\mathcal{R}^{c}_{j}$' after the $(n+1)$th collision, which means the trace of system and all the ancillas interacted with the system except the $j$th ancilla of each reservoir degrees of freedom. And $D(\rho_{1}\parallel\rho_{2})\equiv\textrm{Tr}(\rho_{1}\ln\rho_{1})-\textrm{Tr}(\rho_{1}\ln\rho_{2})$ is the quantum relative entropy between two density matrices $\rho_{1}$ and $\rho_{2}$.

Note that in Ref.~\cite{a6}, it has been shown that if the surrounding of an open quantum system $\mathcal{S}$ are at temperature $\beta^{-1}$, the total entropy flow is $\beta\Delta Q$, where $\Delta Q$ is the heat transfered from $\mathcal{S}$ to its surrounding, i.e., the change of energy of $\mathcal{S}$ due to the interaction with its surrounding. In the present paper of ours, if the temperatures of the two reservoirs are the same and in the thermal equilibrium $\eta^{hc}_{j}=\frac{1}{Z}e^{-\beta\hat{H}^{h}_{j}}\otimes\frac{1}{Z}e^{-\beta\hat{H}^{c}_{j}}$, the second term in Eq. (16),
\begin{equation}
\Delta\mathrm{S}^{re}=\sum^{n+1}_{j=1}\textrm{Tr}_{j}(\tilde{\eta}^{hc}_{j}-\eta^{hc}_{j})\ln\eta^{hc}_{j},
\end{equation}
can be written in the standard thermodynamics form
\begin{equation}
\Delta\mathrm{S}^{re}=-\sum^{n+1}_{j=1}\beta\Delta Q^{\mathcal{\mathcal{R}}}_{j},
\end{equation}
where $\Delta Q^{\mathcal{\mathcal{R}}}_{j}=Tr_{j}[(\hat{H}^{h}_{j}+\hat{H}^{c}_{j})(\tilde{\eta}^{hc}_{j}-\eta^{hc}_{j})]$, is the change of energy of the two reservoirs during the $j$th collision, here $(\hat{H}^{h}_{j}+\hat{H}^{c}_{j})$ represents the total Hamiltonian of `$\mathcal{R}^{h}_{j}+\mathcal{R}^{c}_{j}$'. As we focus on energy-conserving system-reservoir interactions, hence $\Delta Q^{\mathcal{\mathcal{R}}}_{j}=-\Delta Q^{\mathcal{S}}_{j}$ ($\Delta Q^{\mathcal{S}}_{j}$ is the change of energy of the system $\mathcal{S}$). Therefore Eq. (18) can also be written as
\begin{equation}
\Delta\mathrm{S}^{re}=\sum^{n+1}_{j=1}\beta\Delta Q^{\mathcal{S}}_{j},
\end{equation}
and Eq. (19) is the total entropy flow after the $(n+1)$th collision, and accordingly the first term in Eq. (16)
\begin{equation}
\Delta\mathrm{S}^{ir}=D[\rho^{tot}_{n+1}\parallel\rho^{\mathcal{S}}_{n+1}\otimes^{n+1}_{j=1}\eta_{j}^{hc}],
\end{equation}
is the total entropy production after the $(n+1)$th collision. Since
the relative entropy is positive, Eq. (20) indicates the positivity
of the entropy production, i.e., $\Delta \mathrm{S}^{ir}\geqslant0$
(equal to zero only when $\rho^{tot}_{n+1}$ and
$\rho^{\mathcal{S}}_{n+1}\otimes^{n+1}_{j=1}\eta_{j}^{hc}$ are
identical). And in Ref.~\cite{a5,a6,a7} it has been claimed that in this
case (the environment is in the thermal equilibrium) the second law
is fulfilled. However, for the two reservoirs with coherence, i.e.,
Eq. (13) ($p\neq0$), the entropy change for the second term in Eq.
(16), cannot be associated with the heat flow.

In order to study the relation of entropy change and heat exchange with single reservoir ($\mathcal{R}^{h}$ or $\mathcal{R}^{c}$),
firstly we take the system $\mathcal{S}$ and reservoir $\mathcal{R}^{c}$ as a composite system $\mathcal{S}\mathcal{R}^{c}$, and the
change in the von Neuman entropy of $\mathcal{S}\mathcal{R}^{c}$, after the $(n+1)$th collision, can be expressed as
\begin{equation}
\Delta\mathrm{S}_{1}
=D[\rho^{tot}_{n+1}\parallel\rho^{\mathcal{S}\mathcal{R}^{c}}_{n+1}\otimes^{n+1}_{j=1}\eta_{j}^{h}]
+\sum^{n+1}_{j=1}\textrm{Tr}_{j}(\tilde{\eta}^{h}_{j}-\eta^{h}_{j})\ln\eta^{h}_{j},
\end{equation}
where
$\rho^{\mathcal{S}\mathcal{R}^{c}}_{n+1}=\textrm{Tr}_{\mathcal{R}^{h}}(\rho^{tot}_{n+1})$
means the trace of reservoir $\mathcal{R}^{h}$ degrees of freedom.
And we name the second term in Eq. (21),
\begin{equation}
\Delta\mathrm{S}_{1}^{re}=
\sum^{n+1}_{j=1}\textrm{Tr}_{j}(\tilde{\eta}^{h}_{j}-\eta^{h}_{j})\ln\eta^{h}_{j},
\end{equation}
the entropy exchanged with the reservoir $\mathcal{R}^{h}$. Note that if the reservoir $\mathcal{R}^{h}$ is in the
thermal equilibrium $\eta^{h}_{j}=\rho_{\beta}=\frac{1}{Z}e^{-\beta_{h}\hat{H}^{h}_{j}}$, Eq. (22) can be written in the standard thermodynamic form
\begin{equation}
\Delta\mathrm{S}_{1}^{re}=-\sum^{n+1}_{j=1}\beta_{h}\Delta Q^{h}_{j},
\end{equation}
where $\Delta Q^{h}_{j}$ is the change of energy of the reservoir $\mathcal{R}^{h}$ during the $j$th collision, i.e., the heat
flowing from reservoir $\mathcal{R}^{h}$. However, for the state of reservoir $\mathcal{R}^{h}$ with coherence, i.e., Eq. (13) ($p\neq0$),
it can be seen from Eq. (22) that Eq. (23) is not valid. In other words, for the reservoir with coherence, there is no longer a
direct connection between the entropy change of the system and the heat flow from its environment. And Nejad \emph{et.al.} have pointed
out that for the general quantum setting there may be not a direct association between heat flux and entropy change~\cite{a7}.

Similarly, the change in the von Neuman entropy of the composite system $\mathcal{S}\mathcal{R}^{h}$, after the $(n+1)$th collision, can
be expressed as
\begin{equation}
\Delta\mathrm{S}_{2}
=D[\rho^{tot}_{n+1}\parallel\rho^{\mathcal{S}\mathcal{R}^{h}}_{n+1}\otimes^{n+1}_{j=1}\eta_{j}^{c}]
+\sum^{n+1}_{j=1}\textrm{Tr}_{j}(\tilde{\eta}^{c}_{j}-\eta^{c}_{j})\ln\eta^{c}_{j},
\end{equation}
where $\rho^{\mathcal{S}\mathcal{R}^{h}}_{n+1}=\textrm{Tr}_{\mathcal{R}^{c}}(\rho^{tot}_{n+1})$ means the trace of reservoir $\mathcal{R}^{c}$ degrees of freedom. And the second term in Eq. (24), i.e., the entropy exchanged with the reservoir $\mathcal{R}^{c}$,
\begin{equation}
\Delta\mathrm{S}_{2}^{re}=
\sum^{n+1}_{j=1}\textrm{Tr}_{j}(\tilde{\eta}^{c}_{j}-\eta^{c}_{j})\ln\eta^{c}_{j},
\end{equation}
can also be written in the standard thermodynamic form for the thermal equilibrium state of reservoir $\mathcal{R}^{c}$
\begin{equation}
\Delta\mathrm{S}_{2}^{re}=-\sum^{n+1}_{j=1}\beta_{c}\Delta Q^{c}_{j},
\end{equation}
where the heat flowing from reservoir $\mathcal{R}^{c}$ during the $j$th collision is $\Delta Q^{c}_{j}$. Of course, for the state of reservoir $\mathcal{R}^{c}$ with coherence Eq. (26) is also not valid. From Eqs. (17), (22) and (25), we find that the relation $\Delta \mathrm{S}^{re}=\Delta\mathrm{S}_{1}^{re}+\Delta\mathrm{S}_{2}^{re}$ is always satisfied for the two reservoirs being in state (13), which is independent of the value of $p$ and the phase difference of two reservoirs. And clearly, for the two reservoirs are in the thermal equilibrium with the same temperature ($\beta_{h}=\beta_{c}=\beta$), this relation can be written as
\begin{equation}
\sum^{n+1}_{j=1}\Delta Q^{\mathcal{\mathcal{R}}}_{j}=\sum^{n+1}_{j=1}(\Delta Q^{h}_{j}+\Delta Q^{c}_{j}),
\end{equation}
which can be interpreted as that, after the $(n+1)$th collision, the total heat flow (energy change) of the two reservoirs equals to the sum of
the heat flow of the two reservoirs respectively.

\section{Thermal conductance}\label{Sec5}
Thus far, we have only concerned the heat transport with finite
effective temperature difference between the two reservoirs
($T_{h}=2T_{c}=2$). Now we consider the case of small effective
temperature difference between the two reservoirs, and we begin this
study with Fourier's law of heat conduction. Fourier's law of heat
conduction states that the heat current through a classical
macroscopic object is proportional to the applied temperature
gradient~\cite{a8},
\begin{equation}
J=-\kappa\nabla T,
\end{equation}
where $\kappa$ is the conductance. In our case we assume that $T_{h}=T+\Delta T/2$ and $T_{c}=T-\Delta T/2$, and write
\begin{equation}
J_{h}=-\kappa\Delta T,
\end{equation}
then the conductance is obtained from $\kappa=-J_{h}/\Delta T$ by taking the limit $\Delta T\rightarrow0$~\cite{d11}.

$\emph{Thermal state.}$ For thermal initial state of the two
reservoirs, in Appendix B we provide an analytic expression of
steady heat current, and we show that the steady heat current can be
written in the form of $J^{steady}_{h}=-\kappa\Delta T$, therefore the
conductance $\kappa$ is a constant for fixed $T,\delta$ and
$\gamma$. And from Eq. (B1) in Appendix B, we find that $\kappa$
increases with the increase of $T$ at low temperatures
$T\lesssim0.45$; and $\kappa$ decreases with the increase of $T$ at
high temperatures $T\gtrsim0.45$. This indicates that a high
conductance can be obtained at low temperature of reservoir. Note
that a similar result has been obtained in Ref.~\cite{d11} that the
conductance firstly increases (low temperatures) and then decreases
(high temperatures) with the increase of temperature, and at high
temperatures the conductance is proportional to the inverse of
temperature.

\begin{figure}[h]
\includegraphics[scale=0.58]{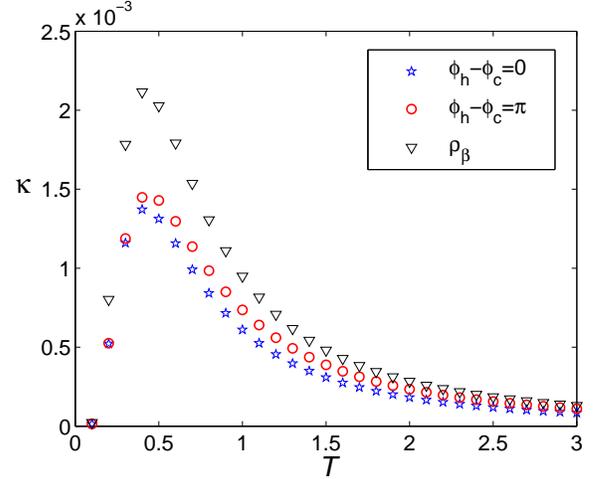}
\centering \caption{(Color online) The conductance as a function of
$T$ for state (13) with $p=0.8$, $\phi_{h}-\phi_{c}=\{0,\pi\}$; and $p=0$, i.e., thermal
state $\rho_{\beta}$. $\gamma=\pi/32$, $\delta=\pi/4$, $\omega=1$, $T_{h}=T+\Delta T/2$ and $T_{c}=T-\Delta T/2$.} \label{Fig4}
\end{figure}

$\emph{State with coherence.}$ From our study, we find that in the
case of $\Delta T\rightarrow0$, generally the steady heat current
can be expressed as
\begin{equation}
J^{steady}_{h}=-\lambda\Delta T+c,
\end{equation}
where $\lambda$ and $c$ are constant for fixed parameters ($p,\gamma,\delta$ and $T$). When the phase difference between the two
reservoirs is $0$ or $\pi$ with fixed parameters ($p=0.8,\gamma=\pi/32,\delta=\pi/4$ and $T=1$), the steady heat current is
\begin{equation}
J^{steady}_{h}=
\begin{cases}
-6.1063\times10^{-4}\Delta T, \ \ \ (\phi_{h}-\phi_{c}=0) \\
-7.3643\times10^{-4}\Delta T, \ \ \ (\phi_{h}-\phi_{c}=\pi) \\
\end{cases}
\end{equation}
and clearly for $\phi_{h}-\phi_{c}=0,\pi$ the constant $c$ in Eq.
(30) is zero, i.e., we can also obtain the conductance now similar
to the case of thermal state. Obviously, from Eq. (31), the
conductance is different for the phase difference
$\phi_{h}-\phi_{c}=0,\pi$. And in order to compare the conductance
for the phase difference $\phi_{h}-\phi_{c}=0,\pi$ and the thermal
state, in Fig. 4, we plot the conductance as a function of
$T$ for $\phi_{h}-\phi_{c}=0,\pi$ and thermal
state $\rho_{\beta}$. It can be seen from Fig. 4 that the relation
$\kappa(\phi_{h}-\phi_{c}=0)<\kappa(\phi_{h}-\phi_{c}=\pi)<\kappa(\rho_{\beta})$
is always true for arbitrary $T$. And they have the
similar behaviors that $\kappa$ firstly increases (at low
temperatures $T\lesssim0.45$) and then decreases (at high
temperatures $T\gtrsim0.45$) with the increase of temperatures, and as mentioned above this result is consistent with
Ref.~\cite{d11}. Physically, this can be easily understood as following.
At high temperature, increasing $T$ would weaken the influence of temperature
difference $\Delta T$ on two reservoirs and lead to the decrease of heat flows.

However, if the phase difference $\phi_{h}-\phi_{c}\neq0,\pi$, the constant $c\neq0$ in Eq. (30), for example when $\phi_{h}-\phi_{c}=\pi/4,5\pi/4$ with the fixed parameters ($p=0.8,\gamma=\pi/32,\delta=\pi/4$ and $T=1$), the steady heat currents are
\begin{widetext}
\begin{equation}
J^{steady}_{h}=
\begin{cases}
-6.3147\times10^{-4}\Delta T-8.8067\times10^{-3},\ \ \ (\phi_{h}-\phi_{c}=\frac{\pi}{4}) \\
-7.3239\times10^{-4}\Delta T+1.1089\times10^{-2}.\ \ \ (\phi_{h}-\phi_{c}=\frac{5\pi}{4}) \\
\end{cases}
\end{equation}
\end{widetext}
From Eq. (32), clearly for $\Delta T\rightarrow0$,
$J^{steady}_{h}<0$ for $\phi_{h}-\phi_{c}=\pi/4$ and
$J^{steady}_{h}>0$ for $\phi_{h}-\phi_{c}=5\pi/4$. In other words,
the steady heat current flows from the reservoir $\mathcal{R}^{h}$
to $\mathcal{R}^{c}$ for $\phi_{h}-\phi_{c}=\pi/4$, while for
$\phi_{h}-\phi_{c}=5\pi/4$ the direction of the steady heat current
is opposite, i.e., from  $\mathcal{R}^{c}$ to $\mathcal{R}^{h}$. In
Fig. 5, we show the steady heat currents as a function of $\Delta T$
and $T$ for initial state (13) with $p=0.8$;
$\phi_{h}-\phi_{c}=\pi/4$ and $\phi_{h}-\phi_{c}=5\pi/4$ in Fig. 5
(a) and Fig. 5 (b), respectively. It can be seen from Fig. 5 that
the steady heat currents increase with the increase of $T$, which is
in contrast to the cases of thermal state and the state with
coherence $\phi_{h}-\phi_{c}=0,\pi$ (Fig. 4) that $J^{steady}_{h}$
decreases with the increase of $T$ at high effective temperatures. This could
appear to be counterintuitive at first, as one always expects that
high temperature $T$ would weaken the effect of
temperature difference $\Delta T$ on heat transfer and lead to the
decrease of heat flows. However, as mentioned above for the
reservoirs with coherence the steady heat current depends not only
on the reservoir effective temperature but also on the coherence (non-diagonal
part of initial state (13)). And physically this can be understood
as following. From Eq. (13), as the effective temperature $T$
increases, the values of the off-diagonal elements of Eq. (13)
increases, i.e., the coherence increases. And in the high
effective temperature limit the coherence of two reservoirs reaches the
maximum, and now the values of the diagonal elements of Eq. (13) are
almost the same (both of them equal to $1/2$ approximately). In a
word the higher the effective temperature, the more the coherence of
reservoir and the more the contribution of the reservoir-coherence
on the steady heat current.

\begin{figure}[h]
\centering
\includegraphics[scale=0.58]{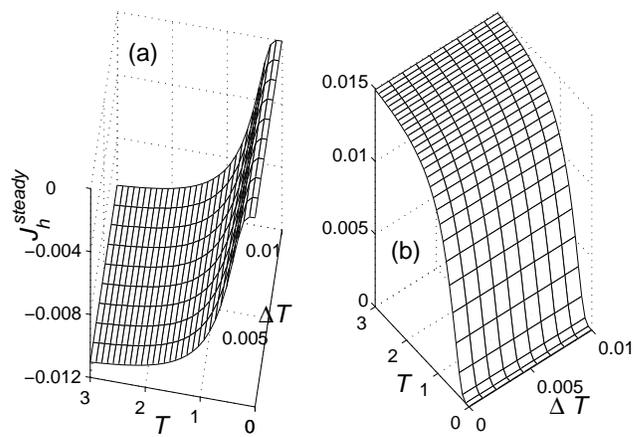}
\caption{Steady heat currents as functions of $\Delta T$ and
$T$ for state (13) with $p=0.8$, $\gamma=\pi/32$, $\delta=\pi/4$;
$T_{h}=T+\Delta T/2$ and $T_{c}=T-\Delta T/2$, and $\omega=1$.
(a) $\phi_{h}-\phi_{c}=\pi/4$, and (b)
$\phi_{h}-\phi_{c}=5\pi/4$.} \label{Fig5}
\end{figure}

We can also understand the result above assisted by the complete
swap case ($\gamma=\pi/2$) as follows. We expand Eq. (14) in series
up to the first order of $\Delta T$ and obtain
\begin{equation}
J^{steady}_{h}=-\lambda_{1}\Delta T+c_{1},
\end{equation}
where
$\lambda_{1}=\frac{1}{4T^{2}}\sech^{2}(\frac{1}{2T})\sin^{2}\delta$
and
$c_{1}=-\frac{1}{4}p^{2}\sech^{2}(\frac{1}{2T})\sin(2\delta)\sin(\phi_{h}-\phi_{c})$.
Clearly, $J^{steady}_{h}$ can be divided into two parts, the first
term in Eq. (33) is the contribution of diagonal elements of state
(13) (thermal state) only. And note that we can also obtain the same
expression of the first term in Eq. (33) from Eq. (B1) in Appendix
B. In other words, for the two reservoirs in thermal states, the
conductance $\kappa=\lambda_{1}$. And the second term in Eq. (33) is
the contribution of all the non-diagonal elements of state (13),
i.e., the total coherence. Now we discuss the effect of phase
difference on the steady heat current in two cases below. (i)
$\phi_{h}-\phi_{c}=0,\pi$. From Eq. (33), when
$\phi_{h}-\phi_{c}=0,\pi$, $J^{steady}_{h}$ returns to the case of
thermal state. However for partial swap ($\gamma\neq\pi/2$), the
first term in Eq. (30) is determined by the diagonal and
non-diagonal elements of state (13) jointly, which leads to the
difference of the conductance for $\phi_{h}-\phi_{c}=0,\pi$ and
thermal state discussed above. (ii) $\phi_{h}-\phi_{c}\neq0,\pi$.
When $\phi_{h}-\phi_{c}\neq0,\pi$, for very small $\Delta T$
($\Delta T\rightarrow0$), the first term in Eq. (33) is negligible,
and Eq. (33) can be written approximatively as
\begin{equation}
J^{steady}_{h}\sim c_{1},
\end{equation}
and obviously for fixed and nonzero parameters $p$, $\delta$ and $\sin(\phi_{h}-\phi_{c})$, Eq. (34) can be re-written as
\begin{equation}
J^{steady}_{h}\sim c_{1}=\xi\sech^{2}(\frac{1}{2T}),
\end{equation}
where $\xi$ is a constant. Because $T>0$ and from Eq. (35), clearly
$J^{steady}_{h}$ increases with the increase of effective temperature $T$, and $J^{steady}_{h}$ is approaching a constant
$\xi$ ($\xi=-\frac{1}{4}p^{2}\sin(2\delta)\sin(\phi_{h}-\phi_{c})$)
in the high effective temperature limit.

\section{Conclusion}\label{Sec6}
In this paper, we have investigated the heat transport between two
nonthermal reservoirs by collision-model-based approach, and we have
studied the effect of reservoir's coherence on the heat current.
Specifically, we have considered a bipartite system consisting of
two identical subsystems, and each subsystem interacts with its
local reservoir, which consists of a large collection of initially
uncorrelated ancillas in a state with coherence. We have realized a
heat transport between two reservoirs by a sequence of pairwise
collisions (inter-subsystem and subsystem-local reservoir). We have
found that the direction of heat current depends on the relative
phases (phase difference between the two reservoirs) strongly. For
example, we have realized heat transfer from the ``cold reservoir''
to the ``hot reservoir'' in the steady regime. This could
appear to be counterintuitive at first, as heat always transfer from
hot reservoir to cold reservoir in general. However it is due to the
contribution of reservoir-coherence on the heat current.

Then we have explored the relation of heat current and entropy
exchanged with the reservoir in our model. We have shown that there
is a linear relation between the heat current and entropy flux for
the reservoir in thermal state, and there is not a direct connection
between them for the reservoir with coherence. Finally, we have
studied  the steady current of heat in the limit of the effective
temperature difference between the two reservoirs $\Delta
T\rightarrow0$. For most of phase differences of two reservoirs, the
steady heat current increases with the increase of effective temperature until to the high effective temperature limit,
and this is in contrast to the thermal states of reservoirs (heat
current decreases with the increase of temperature at high
temperatures). In a word, in the presence of reservoir's coherence
we can observe the effect of reservoir-interference on the heat
transport.

It is noted that in this paper we have used the collision model to
investigate the effect of coherence of reservoirs on the heat flow.
The reason to consider this simple model is that exact solutions can
be obtained for a general class of initial states of reservoirs with
coherence. We expect that some features of the heat flow in this
simple model might be similar to those in more involved but less
tractable heat transfer models so we can gain some insight into the
general feature of effect of reservoirs with coherence on heat flow.

\begin{acknowledgments}
This work is supported by the National Natural Science Foundation of
China (Grant Nos. 11775019 and 11375025), and the Ph.D. research
startup foundation (Grant No. BS201418).
\end{acknowledgments}

\appendix*
\renewcommand{\appendixname}{APPENDIX~\Alph{section}}
\section{}
\renewcommand\theequation{A\arabic{equation} }
After the $(n+1)$th collision, the total state of $\mathcal{S}$ plus
all the ancillas of the two reservoirs which have been interacted with
the system is
\begin{equation}
\rho^{tot}_{n+1}=\hat{U}_{n+1}\cdot\cdot\cdot\hat{U}_{1}\rho^{tot}_{n+1}(0)\hat{U}^{\dagger}_{1}\cdot\cdot\cdot\hat{U}^{\dagger}_{n+1},
\end{equation}
where
$\rho^{tot}_{n+1}(0)=\rho_{0}^{\mathcal{S}}\otimes\prod^{n+1}_{j=1}\eta_{j}^{\mathcal{R}}$
($\eta_{j}^{\mathcal{R}}=\eta_{j}^{h}\otimes\eta_{j}^{c}$), is the
initial state of $\mathcal{S}$ plus the $(n+1)$ ancillas of the two
reservoirs respectively; and
$\hat{U}_{j}=\hat{U}_{\mathcal{S}_{B},\mathcal{R}^{c}_{j}}(\gamma)\hat{U}_{\mathcal{S}_{A},\mathcal{R}^{h}_{j}}(\gamma)\hat{V}_{\mathcal{S}_{A},\mathcal{S}_{B}}(\delta)$
is the unitary operator of the $j$th collision. Hence, the total
energy of the $(n+1)$ ancillas interacted with $\mathcal{S}$ in
reservoir $\mathcal{R}^{h(c)}$ after the $(n+1)$th collision can be
written as
\begin{equation}
\begin{split}
E_{n+1}^{h(c)}=&\mathrm{Tr}[\sum^{n+1}_{i=1}\hat{H}^{h(c)}_{i}\rho^{tot}_{n+1}] \\
=&\sum^{n+1}_{i=1}\mathrm{Tr}[\hat{H}^{h(c)}_{i}\rho^{tot}_{n+1}] \\
=&\sum^{n+1}_{i=1}\mathrm{Tr}[\hat{H}^{h(c)}_{i}(\hat{U}_{n+1}\cdot\cdot\cdot\hat{U}_{1}\rho^{tot}_{n+1}(0)\hat{U}^{\dagger}_{1}\cdot\cdot\cdot\hat{U}^{\dagger}_{n+1})].
\end{split}
\end{equation}
Because $[\hat{H}^{h(c)}_{i},\hat{U}_{j}]=0$ for $j>i$, therefore Eq. (A2) can be written as
\begin{equation}
\begin{split}
E_{n+1}^{h(c)}
=&\sum^{n+1}_{i=1}\mathrm{Tr}[\hat{H}^{h(c)}_{i}(\hat{U}_{i}\cdot\cdot\cdot\hat{U}_{1}\rho^{tot}_{n+1}(0)\hat{U}^{\dagger}_{1}\cdot\cdot\cdot\hat{U}^{\dagger}_{i})] \\
=&\sum^{n+1}_{i=1}\mathrm{Tr}_{i}[\hat{H}^{h(c)}_{i}\tilde{\eta}^{h(c)}_{i}],
\end{split}
\end{equation}
where $\tilde{\eta}^{h(c)}_{i}=\mathrm{Tr}_{\neq
i}^{h(c)}(\rho^{tot}_{n+1})$, means the trace of all except the
$i$th ancilla of the reservoir $\mathcal{R}^{h(c)}$ degrees of
freedom. Clearly, Eq. (A3) is equivalent to the sum of energy change
of the $i$th ancilla (in reservoir $\mathcal{R}^{h(c)}$) during the
$i$th collision.

\section{}
\renewcommand\theequation{B\arabic{equation} }
The expression of steady heat current for thermal state $\rho_{\beta}$ in the limit of small temperature difference $\Delta T$ between the two reservoirs can be obtained,
\begin{widetext}
\begin{equation}
J^{steady}_{h}=\frac{-4e^{\frac{2}{T}}[5+\cos(4\gamma)+4\cos(2\gamma)\cosh^{2}(\frac{1}{2T})+6\cosh(\frac{1}{T})]\sin^{2}\gamma\sin^{2}\delta}{(1+e^{\frac{1}{T}})^{2}T^{2}{\{a(1+e^{\frac{2}{T}})+e^{\frac{1}{T}}[20+7\cos(2\gamma)+4\cos(4\gamma)+\cos(6\gamma)-32\cos^{4}\gamma\cos(2\delta)]}\}}\Delta T,
\end{equation}
\end{widetext}
where $a=11+4\cos(2\gamma)+\cos(4\gamma)-16\cos^{2}\gamma\cos(2\delta)$. Therefore, from Eq. (B1), for thermal state the conductance $\kappa$ can be obtained by the expression $\kappa=-J^{steady}_{h}/\Delta T$. And it is worth mentioning that in the high temperature limit the conductance reduces to
\begin{equation}
\kappa=
\frac{(11+4\cos(2\gamma)+\cos(4\gamma))\sin^{2}\gamma\sin^{2}\delta}{2(3+\cos(2\gamma))(7+\cos(4\gamma)-8\cos^{2}\gamma\cos(2\delta))T^{2}}.
\end{equation}


\end{document}